\documentclass[11pt, a4paper]{article} 

\usepackage{color}
\usepackage{multirow}
\usepackage{tabularx}
\usepackage{caption}

\usepackage[margin=2.5cm]{geometry} 
\usepackage[english]{babel} 
\usepackage[utf8]{inputenc} 
\usepackage{amsmath}		
\usepackage{graphicx}		
\usepackage{pdfpages}
\usepackage{fancyhdr}		
\usepackage{amsfonts,amssymb}
\usepackage{setspace}		
\usepackage{indentfirst}
\usepackage[utf8]{inputenc}
\usepackage{lscape}
\usepackage{longtable}
\usepackage{diagbox}
\usepackage{authblk}

\newcommand{\ignore}[1]{{}}

\begin{document}

\title{Organizing Virtual Conferences through Mirrors:\\The ACM e-Energy 2020 Experience}

\author{Dan Wang$^1$, Arun Vishwanath$^2$, Ramesh Sitaraman$^3$, Iven Mareels$^2$}

\date{\vspace{-0.5cm}
$^1$The Hong Kong Polytechnic University,
$^2$IBM Research, Australia, \\$^3$University of Massachusetts, Amherst
}


	
\setlength{\parindent}{2em} 
\onehalfspacing				

\maketitle

\section{Introduction}
The emergence of the world-wide COVID-19 pandemic has forced academic conferences to be held entirely in a virtual manner. While prior studies have advocated the merits of virtual conferences in terms of energy and cost savings, organizers are increasingly facing the prospect of planning and executing them systematically, in order to deliver a rich conference-attending-experience for all participants.

Starting from March 2020, tens of conferences have been held virtually.
Past conferences have revealed numerous challenges, from budget planning, to selecting the supporting virtual platforms. Among these, two special challenges were identified: 1) how to deliver talks to geo-distributed attendees and 2) how to stimulate social interactions among attendees. These are the two important goals of an academic conference. In this paper, we advocate a {\em mirror program} approach for academic conferences. More specifically, the conference program is executed in multiple parallel (mirrored) programs, so that each mirror program can fit a different time zone. This can effectively address the first challenge.

We present our experience with ACM e-Energy 2020, a small to medium sized conference with 355 registrants. The ACM e-Energy 2020 main program (June 23rd - June 25th, 2020) is single track, with three keynotes, 38 papers in nine technical paper sessions and a poster session. ACM e-Energy 2020 had two mirrors. The peak number of attendees for a session was 120-130 and the average was 70-80, distributed in the two identical mirror programs.

We organize the remaining part of the paper as follows:

\begin{itemize}
\item
In Section \ref{taxmony}, we present one type of categorization of virtual conferences according to how the talks are delivered and how the Q\&A is conducted: talks have four types (off-conference video, offline video, recorded video streaming and live streaming) whereas Q\&A has three types (offline text, live text, and live Q\&A). We argue that a single mirror conference has intrinsic limitation in having live streaming and live Q\&A for all attendees.

\item
In Section \ref{MACS}, we advocate the use of a mirror-based conference program. This allows speakers to present live and attendees to watch the talks live (at a convenient time regardless of their geographical location). Hand crafting the program of a single track conference with 40-50 papers is usually viable. Scheduling a mirror program becomes non-trivial, owing to new constraints on selecting comfortable time slots for the presenters and attendees. Specifically, (1) at least one mirror of the program should fit the comfort period of the attendees, and (2) the program should assign speakers comfortable time slots allowing them to present in all mirrors. We thus formulate and solve a Mirror-program based Academic Conference Scheduling (MACS) problem. The ACM e-Energy 2020 program was hand crafted due to a tight schedule. In this paper, we use MACS to generate a new ACM e-Energy 2020 program and present insights by comparing it with the existing program.

\item
In Section \ref{planning-execution}, we report the planning and execution of ACM e-Energy 2020 conference. We present the objectives, platform choices, personnel roles and responsibilities, and the design of different types of sessions. We present our choices so as to eliminate any hiccups with respect to the planning and conference organization.

\item
In Section \ref{survey-results}, we present the feedback from ACM e-Energy 2020 attendees. We conducted a survey and collected valuable experience from nearly 100 attendees. The overall acceptance of a mirror program was overwhelming. 

\item
In Section \ref{conclusion}, we present lessons learned and conclude the paper.

\end{itemize}

\subsection{Related Work}

With the outbreak of COVID-19, teaching, business meetings, academic conferences, etc., are all going virtual and guidelines are being developed to render them effectively \cite{Barral20}\cite{Price20}. Recently, ACM has established a special task force to draft guidelines for holding ACM conferences virtually, see \cite{acm20}.

One of the first reports on a virtual academic conference was ASPLOS 2020 \cite{Larus20}. The paper presented how the decision was made to move ASPLOS 2020 virtual and how ASPLOS 2020 was managed. The registration fee was waived for the virtual event; and it reports that the number of registrants was higher than in past years. Follow up experience were presented by ICPE 2020 \cite{Iosup20} and PAM 2020 \cite{Misa20}. ICPE 2020 organizers stated that flexilibity was important \cite{Iosup20}. PAM 2020 presented a carefully designed questionnaire \cite{Hohlfeld20}, including prior and post conference questionnaires. We adopted some good questions from PAM 2020. Both ICPE 2020 and PAM 2020 noticed the difficulty for geo-distributed attendees to attend talks. ICPE 2020 chose to set their conference duration to 3 hours per day to increase flexibility. To make this possible, the in-conference talk was two minutes, and each paper had a 20 minute video that the attendees could watch offline. PAM 2020 chose to follow the time of Alberta, Canada, as their conference time, primarily because this was the original conference location and the attendees are usually located in North America.

Other conferences accumulated experience from different aspects. For example, PerCom 2020 and IoTDI 2020 developed the guideline for using Zoom as the supporting platform. WCNC 2020 and INFOCOM 2020 used web video broadcasting in addition to Zoom. This solved the problem for attendees with restriction in accessing tools like Zoom.

These past experiences have no doubt helped the organization of ACM e-Energy 2020. In this paper, we contribute a design of a mirror program approach to address the problem of serving geo-distributed attendees effectively, and discuss the lessons we learned from its implementation.

\ignore{
\begin{figure}[!h]
	\includegraphics[width=13cm]{img/Compare_Figure.png}
	\caption{Comparison of Virtual Academic Conferences}
	\label{figure:conf_compare-2}
\end{figure}
}


\section{Categorization of Virtual Conferences}
\label{taxmony}

Different conferences have different objectives, scale, and restriction. In this paper, we present one categorization along two dimensions - paper talk types and Q\&A types, and examine how synchronous the talks and Q\&A were for the attendees.

Paper talks can be classified into four types: 1) off-conference video, i.e., all talks are uploaded to an online site but there is no concrete conference schedule for paper talks; 2) offline video streaming, there is a conference schedule and attendees can watch video talks during the conference; 3) recorded video streaming, paper talks are scheduled, yet only recorded videos are broadcasted; and 4) live video streaming.
Q\&A can be classified into three types: 1) offline Q\&A, 2) live text Q\&A and 3) live Q\&A.

Note that these types are progressive, i.e., live video streaming means that recorded video, offline video and off-conference video can also be supported but not vice versa. Similarly, live Q\&A means that live text and offline Q\&A can also be supported, but not vise versa.

Live video streaming along with live Q\&A better emulate a physical conference. Different conferences, however, have to balance different requirements and may not adopt the same approaches. For example, WCNC 2020 is a multi-track conference with a large number of attendees. To serve the large number of attendees, it chose the Zoom webinar mode, which broadcasts video talks through the web. Since there is no concept of ``meeting room" for attendees to enter, only live text Q\&A can be supported.


Most conferences to date are single mirror conferences (not to be confused with single track conference). Clearly, a fraction of the geo-distributed attendees will miss out on live presentations and have to be content with watching offline videos. The only conference with multiple mirrors prior to ACM e-Energy 2020 was CVPR 2020. CVPR 2020 had two mirrors but did not have live video streaming. This may be because that CVPR had 1470 papers and the talks are short (a CVPR oral presentation is only 5 minutes). The switching overhead per paper can be high in such circumstance.

\section{Mirror-program based Academic Conference Scheduling (MACS)}
\label{MACS}

In this section, we will present our approach to systematically design a mirror-based conference program. We     begin by describing the motivation behind creating such a program.

\subsection{ACM e-Energy 2020 Program and the Motivation for MACS}

ACM e-Energy 2020 is single track, with 38 academic papers, three keynotes, and 14 posters. The program\footnote{https://energy.acm.org/conferences/eenergy/2020/program.php} was hand crafted. At the time of developing the program, we had no information on the number and geo-distribution of the attendees.

We initially considered three mirrors, representing North America, Europe and Asia Pacific, but decided against this to reduce segregation, i.e., to prevent a mirror from having very few participants. We also realized that due to time constraints, hosting three mirrors would have led to considerable overhead for the organizers. So we chose two mirrors - a London time (BST) mirror and an LA time (PDT) mirror. The rationale was to space the time zones of the mirrors far apart to increase coverage.

An initial three-day program was first developed and then copied to both mirrors. This initial design was similar to a physical conference. 
We made an adjustment for the keynote sessions to ensure keynote speakers present live in the early mirror, i.e., the London mirror.

By observing the program, in hindsight, we see that it only coarsely accounts for the convenience of participants. There is much that can be done to accommodate more friendly times for the attendees and speakers. Such fine-granularity scheduling is beyond what can be hand crafted, motivating the need to develop a computer aided scheduling program.

\subsection{Problem Formulation}

We now briefly present the design philosophy of the MACS problem.
An academic conference program has three components: attendees, sessions, and speakers. Note that we use sessions instead of papers. This is because session development usually is unique from conference to conference: not only the paper sessions differ greatly but also a conference has non-paper sessions. A paper-into-session scheduling is a separate problem that is independent to creating a mirror program. We argue that the PC chairs can first develop the sessions, e.g., paper sessions, keynote sessions, poster sessions, award sessions, etc. These sessions can then become the inputs for the MACS program.

\begin{itemize}
\item
\textbf{Attendees:} An attendee has three attributes: 1) a time zone, e.g., EDT, 2) a comfort period, e.g., 8:00am - 8:00pm,
and 3) presence ratio, the percentage of the conference time that falls in the comfort period. We say that the attendee is {\em satisfied} if the presence ratio is greater than a threshold $\Delta$. If $\Delta = 100\%$, this means that this attendee can attend all sessions live.

\item
\textbf{Sessions:} Sessions have two attributes: 1) length, e.g., 2 hours (with 1.5 hours of paper talks and 30 minutes of break) and 2) a set of speakers associated to a session.

\item
\textbf{Speakers:} A speaker is an attendee with two additional attributes: 1) the associated sessions, and 2) a presentation comfort period. This period differs to the attendee comfort period. This is because a presentation is short and speakers usually have an obligation to deliver their presentation. Therefore, the presentation comfort period could be longer than the comfort period of the attendees. Speakers are {\em satisfied} if their talk sessions are assigned to their presentation comfort period.
\end{itemize}

A {\em conference schedule} is a number of consecutive sessions with a uniform session order across different mirrors, e.g., the power grid session of ACM e-Energy 2020, if assigned to 3:00pm - 4:30pm in a mirror, should remain 3:00pm - 4:30pm in all other mirrors. A good conference schedule needs to maximize its 1) {\em attendance ratio}, the ratio of attendees that are satisfied, and 2) {\em the speaker presence ratio}, the ratio of speakers that are satisfied.

\textbf{The MACS problem:} Given a set of attendees, a set of sessions, a set of speakers, the number of mirrors (e.g., two), find the conference schedule, so that the speakers presence ratio is greater than a threshold $\Omega$ and the attendance ratio is maximized.

\subsection{Evaluation}

\begin{figure}[!h]
	\centering
	\includegraphics[width=15cm]{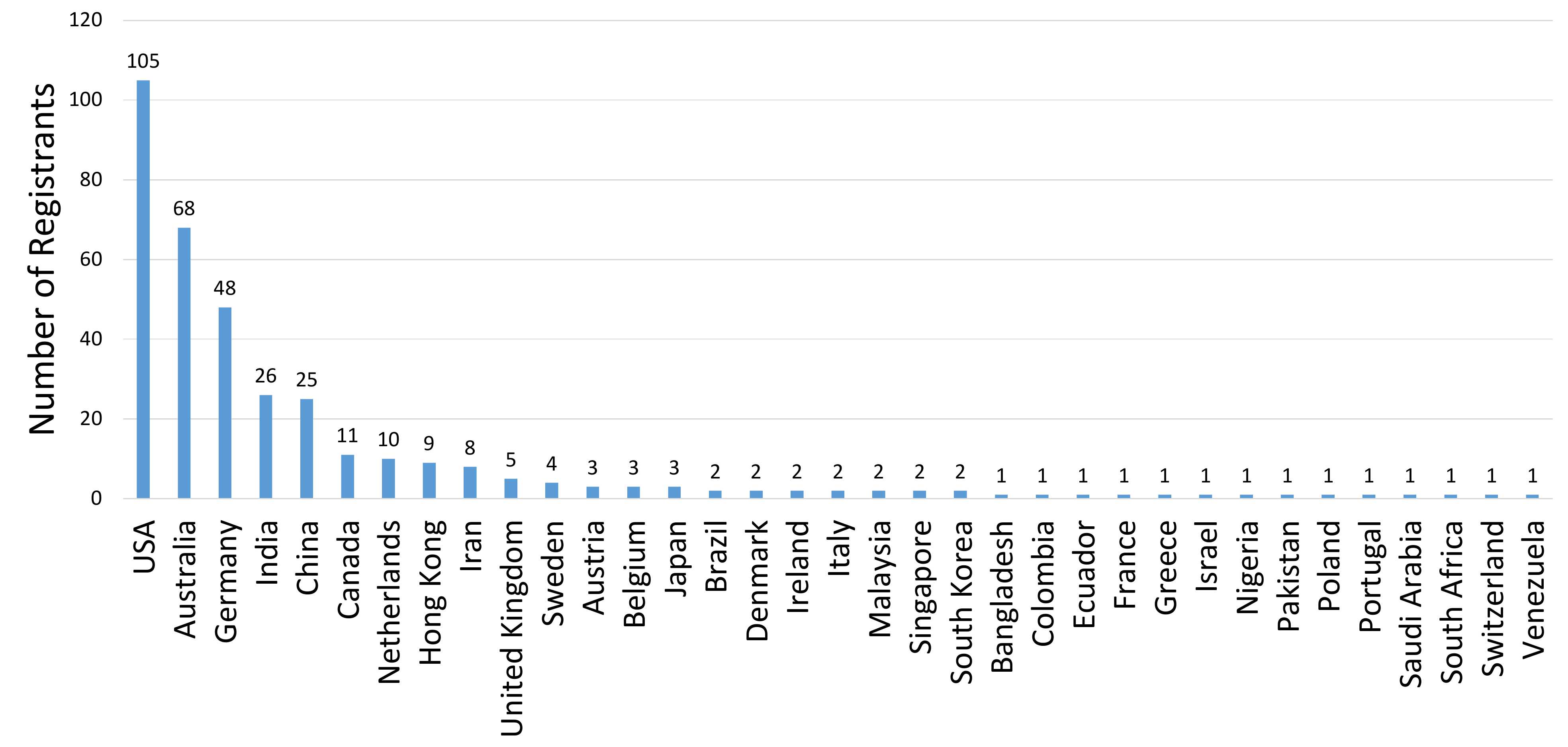}
	\caption{The time zones of ACM e-Energy registrants.}
	\label{figure:registrants}
\end{figure}

We developed a heuristic for the MACS problem. We evaluate MACS using the ACM e-Energy 2020 data. There is a total of 355 registrants and their time zones are shown in Fig. \ref{figure:registrants}. The default comfort period of attendees is set to 8:00am - 8:00pm. We use the same sessions as ACM e-Energy 2020. The total conference time for ACM e-Energy 2020 is 27.5 hours. The default presence ratio of the speakers is 0.8.

\begin{figure}[!h]
	\centering
	\includegraphics[width=13cm]{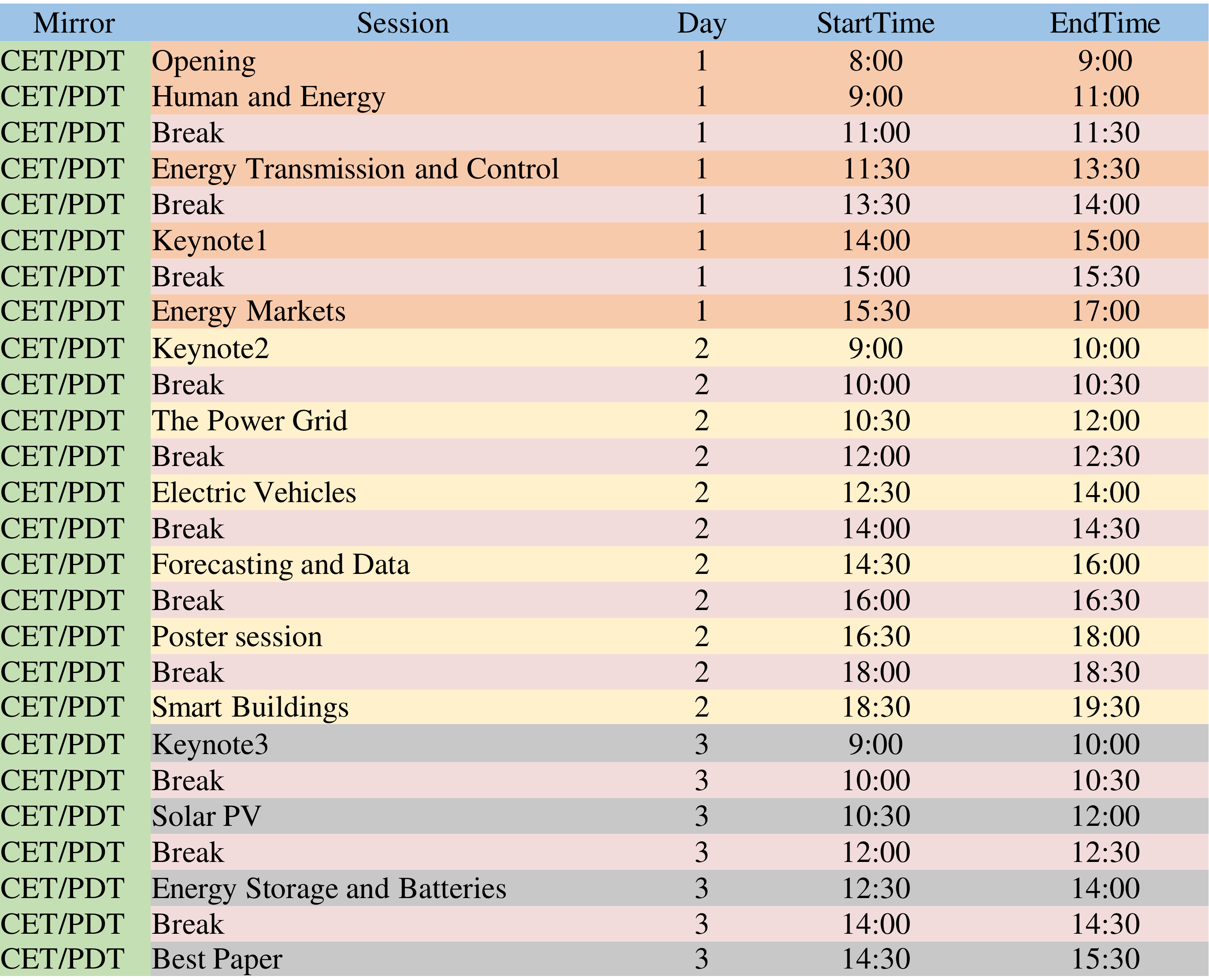}
	\caption{A new time table generated from MACS for ACM e-Energy 2020.}
	\label{figure:timetable}
\end{figure}

\begin{figure}[!h]
	\centering
	\includegraphics[width=13cm]{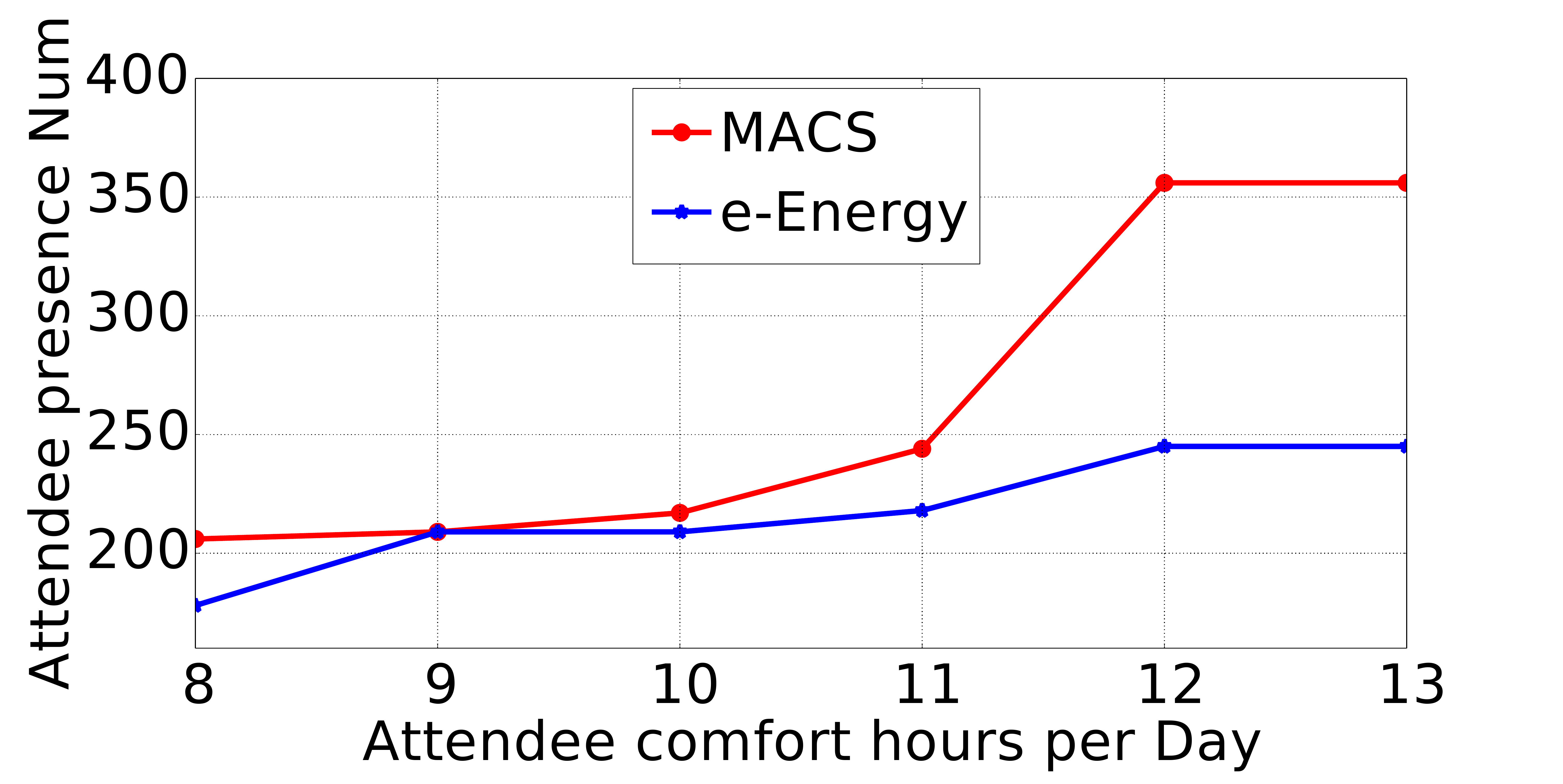}
	\caption{Number of satisfied attendees as a function of comfort period.}
	\label{figure:comforthours}
\end{figure}

Fig. \ref{figure:timetable} show an example of a new ACM e-Energy 2020 program generated from MACS. We note that the session sequence differs from ACM e-Energy 2020. We also note that the time zones of the two mirrors are CET (Central European Time) and PDT instead of London Time and PDT. This reflects that a slight switch from London time to CET can cover more Asia-Pacific attendees of ACM e-Energy 2020.

Fig. \ref{figure:comforthours} shows the number of satisfied attendees as a function of the comfort period per day, here in x-axis, 8 denotes 8:00 - 16:00, 9 denotes 8:00 - 17:00, etc.
Clearly, the longer the comfort period, the more satisfied attendees there are in a conference. For example, when the comfort period is 10 hours, the number of satisfied attendees is 217; and when the comfort period is 12 (8:00 - 20:00), the number of satisfied attendees increases to 355, i.e., all registrants are satisfied. We can also see that the MACS program outperforms the existing hand crafted ACM e-Energy 2020 program.

\begin{figure}[!h]
	\centering
	\includegraphics[width=13cm]{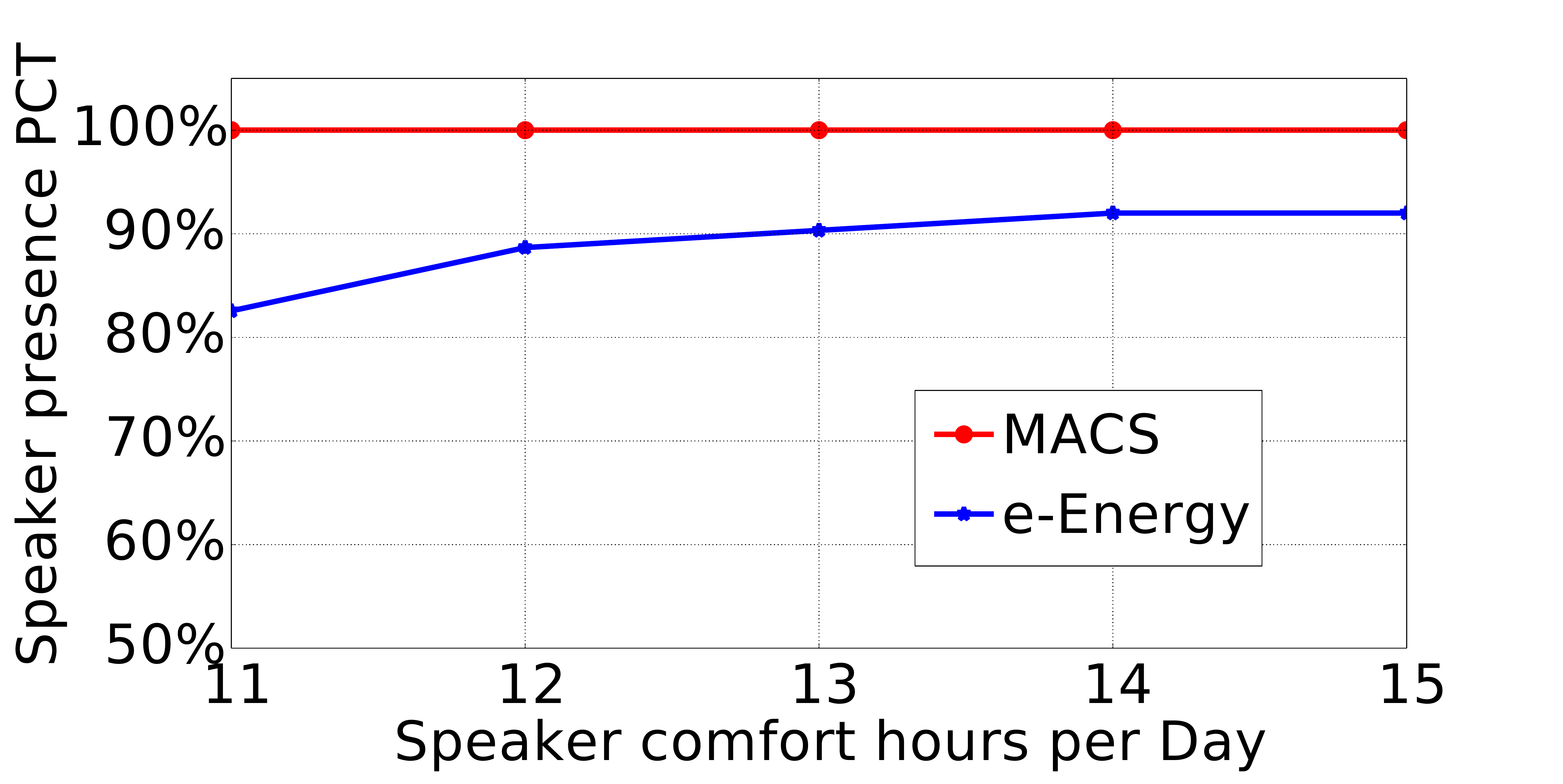}
	\caption{Average proportion of speakers present on live of different comfort period.}
	\label{figure:workinghours}
\end{figure}

Fig. \ref{figure:workinghours} shows the fraction of speakers who present live as a function of the speaker's presentation comfort period per day, here in x-axis, 11 denotes 8:00 - 19:00, 12 denotes 8:00 - 20:00, etc. Obviously, the longer the speaker's comfort period, the larger proportion of speakers can present live. Note that the proportion of speakers present in their comfort period can only reach to 90\% when the speaker's comfort period increases to 15 hours in the existing e-Energy program, and the fraction is always 100\% in MACS, meaning that all speakers are comfortable under the MACS program even when there are only 11 hours in the comfort period.

\begin{figure}[!h]
	\centering
	\includegraphics[width=12cm]{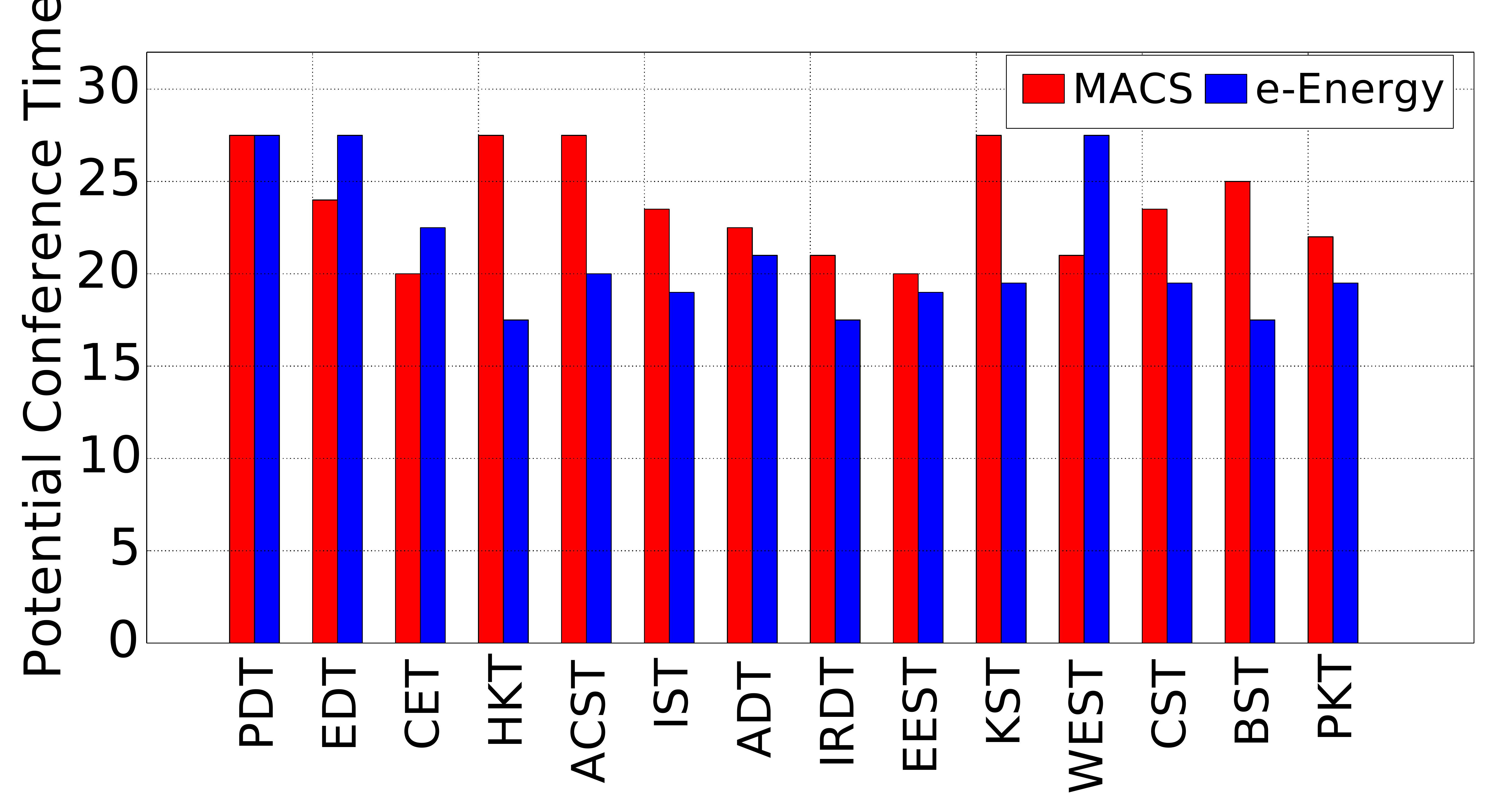}
	\caption{Potential conference time of attendees from different time zones.}
	\label{figure:timezone_average_pct}
\end{figure}

We take an in-depth look at the distribution of the potential conference period of attendees. Fig. \ref{figure:timezone_average_pct} shows the results. We see that MACS, with only a few exception, improves the potential conference time for the attendees in most time zones.

MACS is a multi-objective problem. Conferences can have different weights on different objectives. For example, some may emphasize on the attendance ratio while others may care about the speaker presentation ratio (i.e. they want more speakers to present live), and yet others may want a longer paper presentation time. We argue that there are four main scheduling criteria for the MACS problem: 1) {\em the talk duration of a paper}, e.g., 18 minutes + 4 minutes of Q\&A, 2) {\em the ratio of satisfied attendees}, 3) {\em the presence ratio of an attendee}, e.g., the ratio 0.5 means that 50\% of conference time matches with the comfort period of an attendee, and 4) {\em the ratio of satisfied speakers}, e.g., the ratio 0.6 means that 60\% of speakers can present live in all mirrors in the conference. Clearly, these four criteria conflict with each other. For example, if we want the presence ratio of each attendee to increase, the ratio of satisfied attendees will decrease; if we increase the length of talks, the conference time will increase but the ratio of satisfied attendees, as well as that of the speakers, will decrease.

\begin{figure}[!h]
	\centering
	\includegraphics[width=12cm]{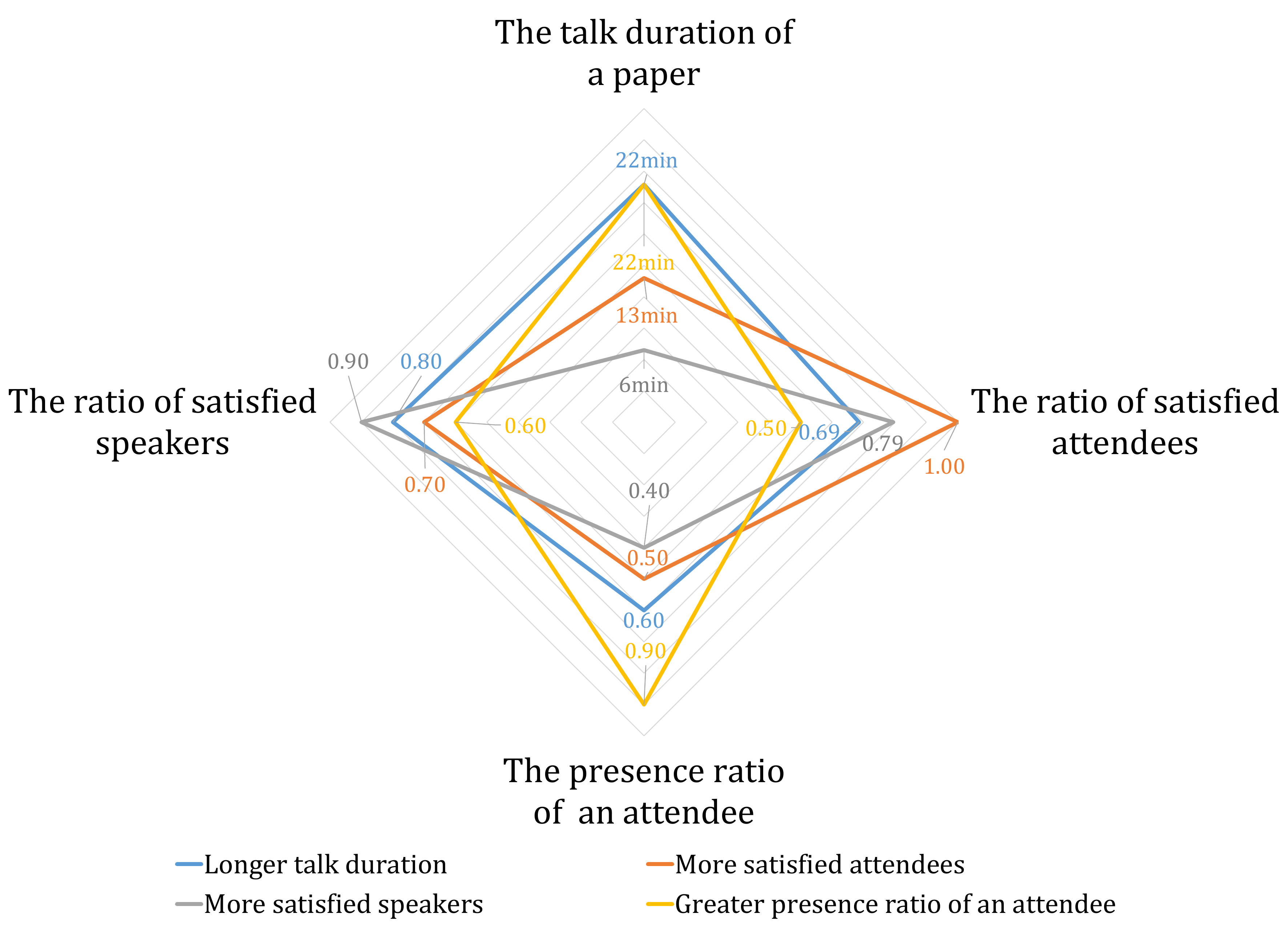}
	\caption{Four main criteria of MACS problem.}
	\label{figure:criteria_macs}
\end{figure}

Fig. \ref{figure:criteria_macs} shows the relationships among the four criteria of the possible choices for ACM e-Energy 2020. There are four vertices in the figure, each corresponding to the above mentioned criteria. We can see that if we want to increase the value of one vertex, we should tune the value of other vertices to fulfil the objective. For example, the orange line in Fig. \ref{figure:criteria_macs} represents the objective to have more satisfied attendees. Maximizing this would entail decreasing the presence ratio of attendees to 0.5, the presence ratio of speakers tuned to 0.7 and the talk duration of papers is fixed to 13 minutes. This leads to 355 satisfied attendees, i.e. all the registrants. The other three cases, demonstrated in blue, grey and yellow line, show the same property. Therefore, the four criteria of MACS problem are interrelated. The currently ACM e-Energy 2020 program is the blue color box.

\section{Conference Planning and Execution}
\label{planning-execution}

We now point to some of the important activities in our quest to host a mirrored conference and outline a planning guide that we hope will enable organisers to run a successful mirrored conference.

\subsection{Planning Objectives}

The call for papers closed in the first week of Feb 2020. At that point, COVID-19 was still in its infancy, and we were hopeful that a physical conference could be held in Melbourne, Australia. As the pandemic began to tighten its grip across countries around the world, it became apparent that we will have to host the conference virtually, else risk being it cancelled altogether. Nevertheless, we did not want to deviate from our original design objectives, namely to give participants the experience of attending a physical conference as much as possible. To do so, we embarked on designing the virtual conference to support the following features: (i) to have each paper presented live in at least one mirror, (ii) to have live Q\&A, (iii) to have live keynote presentations, (iv) to have live poster sessions, and (v) to provide interactive social networking sessions.

\subsection{Platform Choices and Testing}

We used a combination of Zoom and Slack to host the conference. The former was used for video streaming while the latter was used for Q\&A. As our survey results show, these tools were indeed popular with the participants - over 90\% of the respondents had experience with Zoom while nearly 65\% were comfortable with Slack. Keeping things simple and planning the logistics carefully ensured that the conference went smoothly, devoid of any glitches.

\begin{itemize}
\item Choice of Slack: Ironically, we had trialled Slack at ACM e-Energy 2018 as a means to foster increased interaction between the authors and attendees. We had created one Slack channel per paper, resulting in a total of 42 channels. While the uptake was moderate, the feedback was that one channel per paper was a bit too onerous to navigate. Thus, in 2020, we created one channel per paper session, along with one each for the keynotes and posters, resulting in a total of 11 channels. We are glad to report that this arrangement worked favourably with the attendees engaging in numerous productive discussions. Overall, 185 attendees posted in excess of 3000 messages during the conference.

\item Choice of Zoom: As many universities switched to online learning due to COVID-19, we realized that Zoom's popularity within the academic community was growing. We therefore decided to use Zoom for the conference while also recognising that it provides features such as breakout rooms - that can be used to emulate session breaks, hallway conversations and poster sessions - which we felt would enhance the conference experience significantly. Other features that we found useful were - assigning alternate hosts, creating multiple co-hosts, enabling virtual backgrounds and recording in the cloud. In addition, it was available at an attractive price point of AUD 30 per user per month. At the peak, we observed about 120 - 130 unique attendees across the BST and PDT mirrors.

\item Choice of Box: We requested recorded videos from all speakers, both for recorded video streaming and for backup if there are technical problems. We choose Box for storage.
\end{itemize}

We would like to mention that running several dry runs using Zoom eliminated a vast majority of unexpected surprises. There is one point, however, worth mentioning. We discovered that creating Zoom meetings for both the mirrors from a single login became problematic. This meant only one Zoom meeting could be active. In other words, when the PDT meeting started, the BST mirror would automatically disconnect; there was 2 hours overlap between the two mirrors. To overcome this issue, we ensured that the BST and PDT mirror meetings were created by two different Zoom masters with two separate logins. All of the above steps ensured that the conference  - the first with two live mirrors - was held successfully without a glitch.

\subsection{Personnel Roles and Responsibilities}

Besides the conference organizers, the session chairs and Zoom masters greatly facilitate the execution of the conference. We selected a session chair for each session and requested the session chairs to be presence in both mirrors. We ensured that his session in both mirrors fall into the comfort periods of the session chairs.

A few weeks before the conference, we contacted the authors to know (i) who would be presenting the paper, and (ii) in which mirror/mirrors would they be presenting live. We also requested them to upload a recording of their talk in the designated Box folder. This information was shared with the session chairs and zoom masters (see below) in advance, eliminating any uncertainty around how a talk in a session would be delivered.

We recruited student volunteers to serve as `zoom masters'. Their role was to facilitate the use of the Zoom platform, thereby ensuring that all presentations and sessions were executed flawlessly. There was a primary zoom master and a backup zoom master for every session. For the BST mirror, the zoom masters were from Australia (2), Hong Kong (2), Germany (1) and Italy (1). For the PDT mirror, the zoom masters were all from USA (4). Towards that end, they:
\begin{itemize}
\item First, familiarized themselves with Zoom and its features such as how to create breakout rooms, how to record talks, and all the essential functions for their session.
\item Then, they introduced themselves to the respective presenters and tested a dry run of the talks. Recordings were also watched to ensure consistency and quality of video streaming.
\item Finally, they got in touch with the session chairs and the responsibilities of each of them were clearly identified. The session chairs' role was that of a regular conference session chair- introducing the speaker, moderating Q\&A, keeping time. The zoom masters took care of everything else - hosting the meeting and recording sessions.
\end{itemize}

\subsection{Managing Conference Sessions}

Beside the technical paper sessions, we present our experiences of the opening, best paper, keynotes, and poster sessions.

\subsubsection{Opening and Best Paper Sessions}

We opt to conduct these two sessions live in both the mirrors. Although this session was held first in the BST mirror, which was also recorded, it could not be played back in the PDT mirror because Zoom incurs some processing time before it makes the recording available. Moreover, processing of the videos begins only after the meeting concludes; in this case after the BST mirror had ended for the day. Due to this technicality, we had to conduct the sessions live.

It was not particularly onerous since they lasted for about 30 minutes each. Yet we note that the attendance of the opening in both mirrors was high, whereas the attendance of the best paper session differs greatly in the two mirrors - the earlier BST mirror had significantly more. We believe that the announcement of the best paper spread out fast, and the interest in the session immediately decrease when the announcement is out.

\subsubsection{Keynote Sessions}

The three keynotes were presented live in the BST mirror. As with the paper authors, we had also requested the keynote presenters to upload a recording of their talk onto the Box folder. This proved to be very helpful because we could use it to stream their talks in the PDT mirror.  For the same reasons mentioned above, the recording of their live talks (in the BST mirror) was not available in time for playback in the PDT mirror. Our careful planning helped overcome this tricky situation. Finally, Slack was actively used by the attendees to engage in a lively Q\&A with the keynote speakers.

\subsubsection{Breaks and Networking}

Breaks, of 30 minutes, were introduced between sessions for networking opportunities and hallway conversations. There was a `lunch' break of an hour incorporated into the program. We also created dedicated break sessions around lunch and after the end of each day.

The breakout rooms feature in Zoom proved to be very helpful. The zoom master created these rooms with about 6 to 8 people in each breakout room and randomly allocated the attendees to the rooms. As the survey results show, a vast majority of the participants appreciated the use of this feature for increased socializing and interaction.

\subsubsection{Poster Session}

We hosted a poster session with 14 posters. Before the conference, the authors were asked to record a 2 minute presentation and to upload it to the Box folder. The video collection was then streamed during the sessions in both the mirrors. Following this, breakout rooms were created, similar to the above, where each breakout room had one or two authors present to discuss their posters. This time, participants were assigned the `co-host' status, enabling them to freely move between breakout rooms, thus emulating a physical conference where the audience may move between posters as they choose to learn more about the work they find interesting.

\begin{figure}[!h]
	\centering
	\includegraphics[width=12cm]{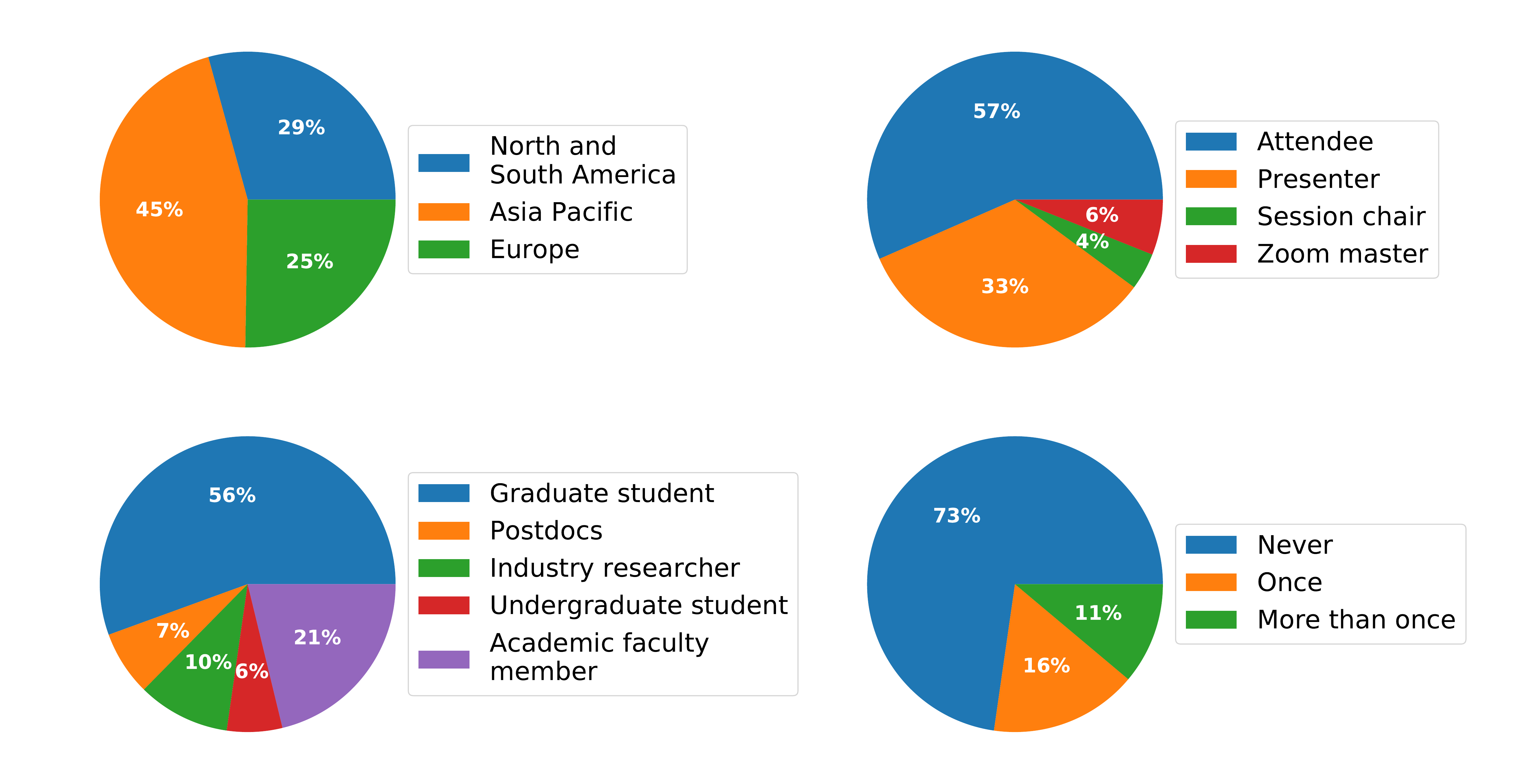}
	\caption{General background of survey participants: (a) Region, (b) Role, (c) Profession, (d) Past experience of ACM e-Energy.}
	\label{figure:participants_info}
\end{figure}

\section{ACM e-Energy 2020: Feedback from Attendees}
\label{survey-results}

We conducted a survey of conference participants by asking them to fill out an on-line questionnaire. 99 participants (28\%) responded to the survey. We summarize the general information of all respondents in Fig. \ref{figure:participants_info}. The distribution of the survey participants generally reflects the distribution of the registrants.

\begin{figure}[!h]
	\centering
	\includegraphics[width=12cm]{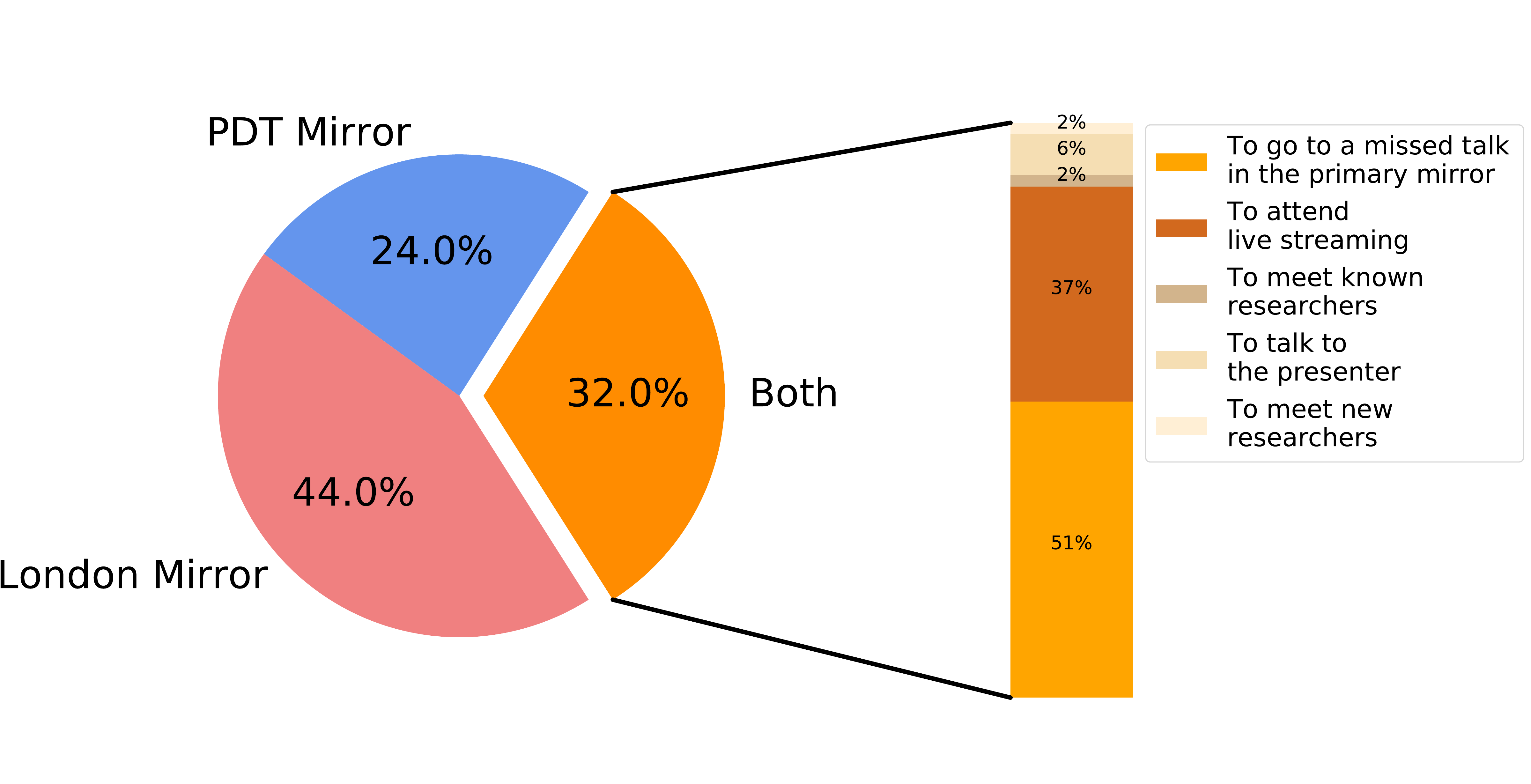}
	\caption{An overview of mirror attendance.}
	\label{figure:mirror_objective}
\end{figure}

ACM e-Energy 2020 was held in two mirrors. We primarily show the experience of participants on the mirror setting. A general information is that in our mirror setting (London time and Los Angeles time), participants from Europe to China (15:00pm - 1:00am) fit best to the London mirror, participants of America fit best to the PDT mirror and participants of Australia may need to attend both mirrors.

\begin{figure}[!h]
	\centering
	\includegraphics[width=12cm]{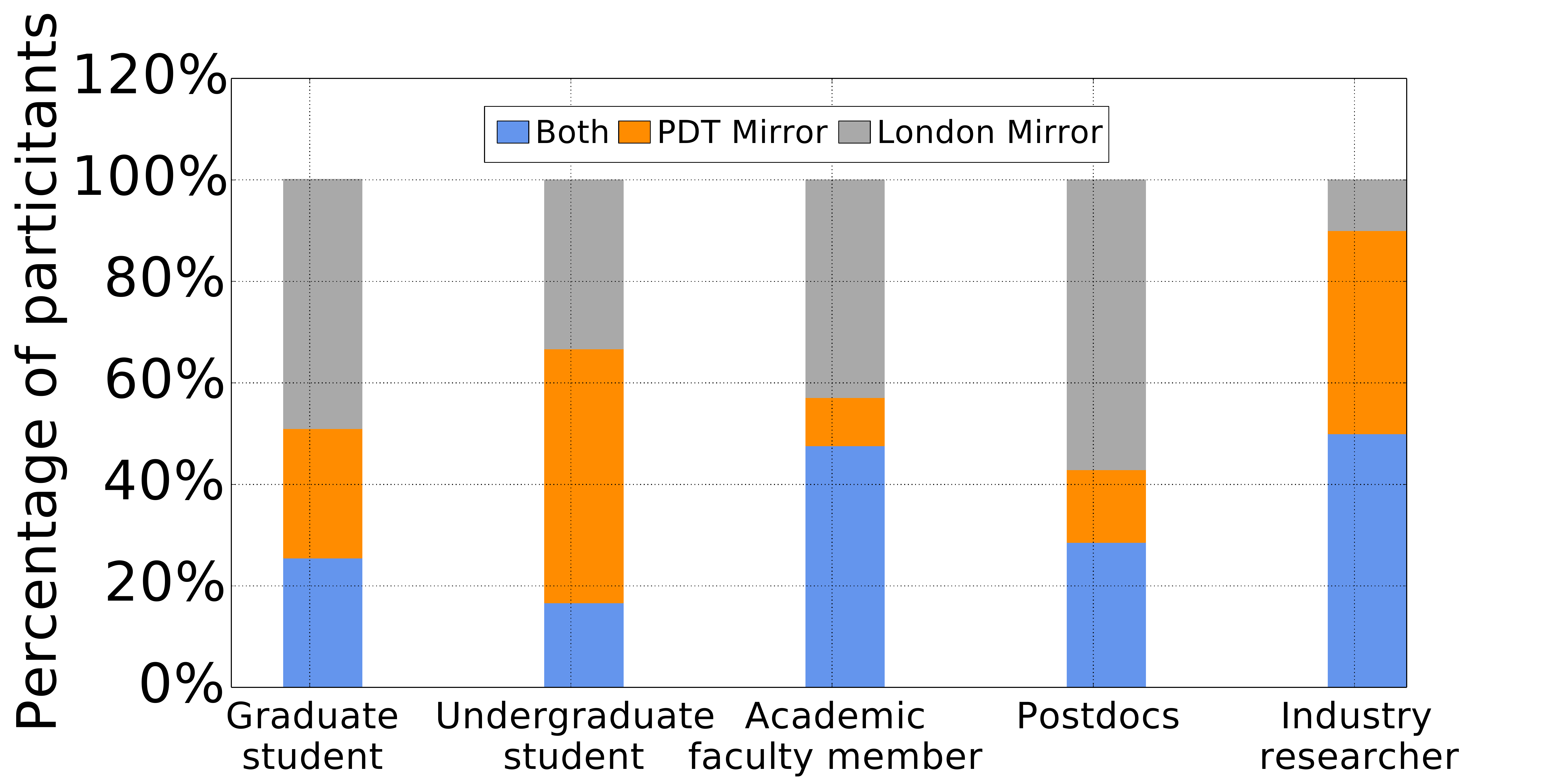}
	\caption{Mirror selection of each profession.}
	\label{figure:mirror_profession}
\end{figure}

In our survey, we asked the participants which mirror they attended. Fig. \ref{figure:mirror_objective} shows the results. We can see that 68\% of participants choose to attend one mirror (24\% to attend the PDT mirror and 44\% to attend the London mirror). 32\% of them attended both. For those who attended both mirrors, we asked their motivation. We see that most of them go to look for talks (51\% to go to a missed talk and 37\% to attend live streaming) and a few search for social contact. This shows that attendees primarily use the mirror program to attend talks.


\begin{figure}[!h]
	\centering
	\includegraphics[width=12cm]{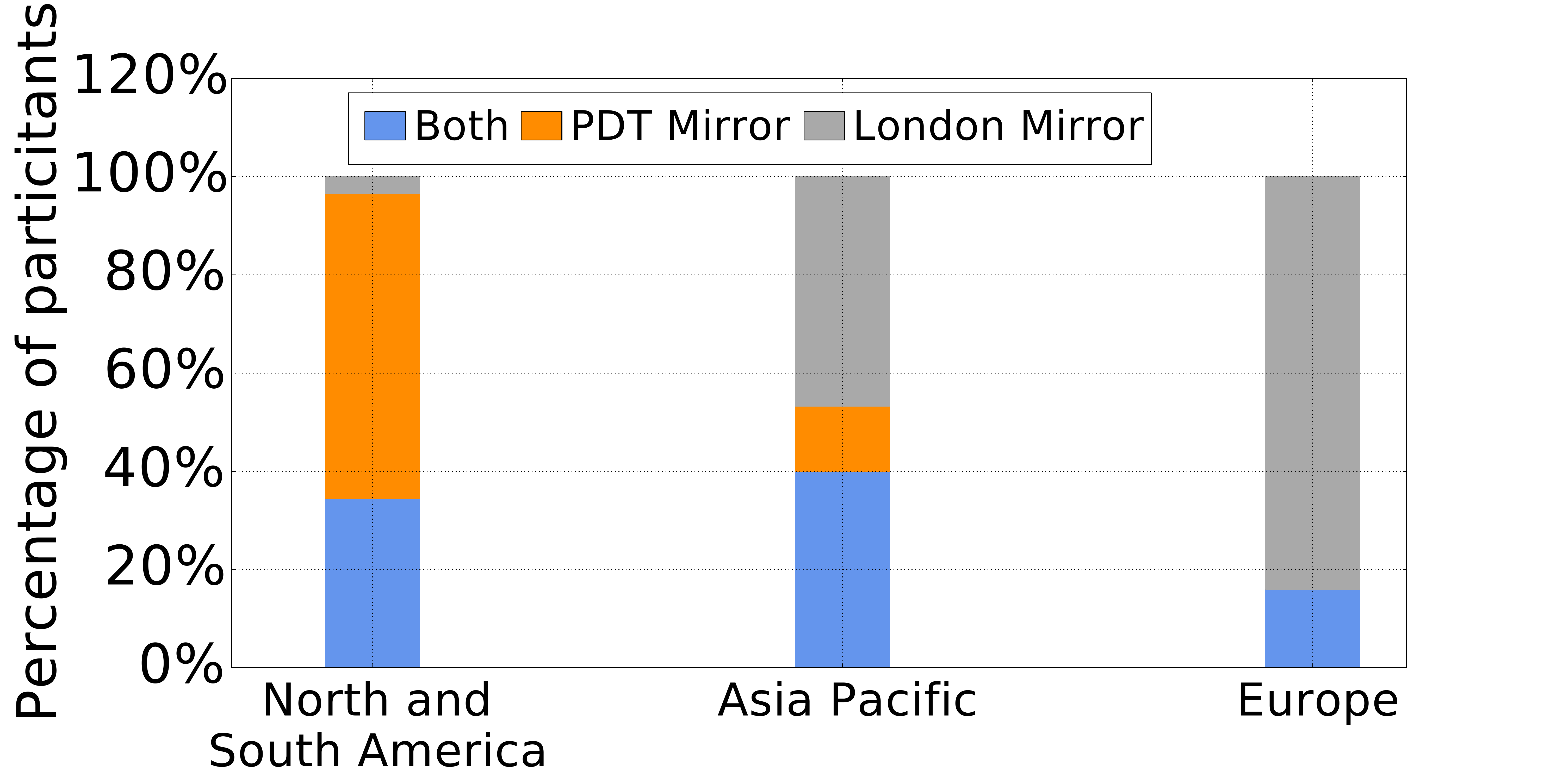}
	\caption{Mirror selection of each region.}
	\label{figure:mirror_region}
\end{figure}

We take an in-depth look at the participants that attended both mirrors. We see in Fig. \ref{figure:mirror_profession}, that academic faculty members and industry researcher are more diligent, 50\% attended both mirrors; whereas 30\% of postdocs and 25\% of graduate students attended both mirrors, and less than 20\% of undergraduate students attended both mirrors.

Fig. \ref{figure:mirror_region} shows who are the attendees of each mirror. More than 80\% of European participants only attended the London mirror and 20\% attended both. As a comparison, there are more American participants attending both mirrors. This may be because the London mirror went online first with live keynotes. Asia-Pacific attendees has the highest ratio to attend both mirrors.

\begin{figure}[!h]
	\centering
	\includegraphics[width=12cm]{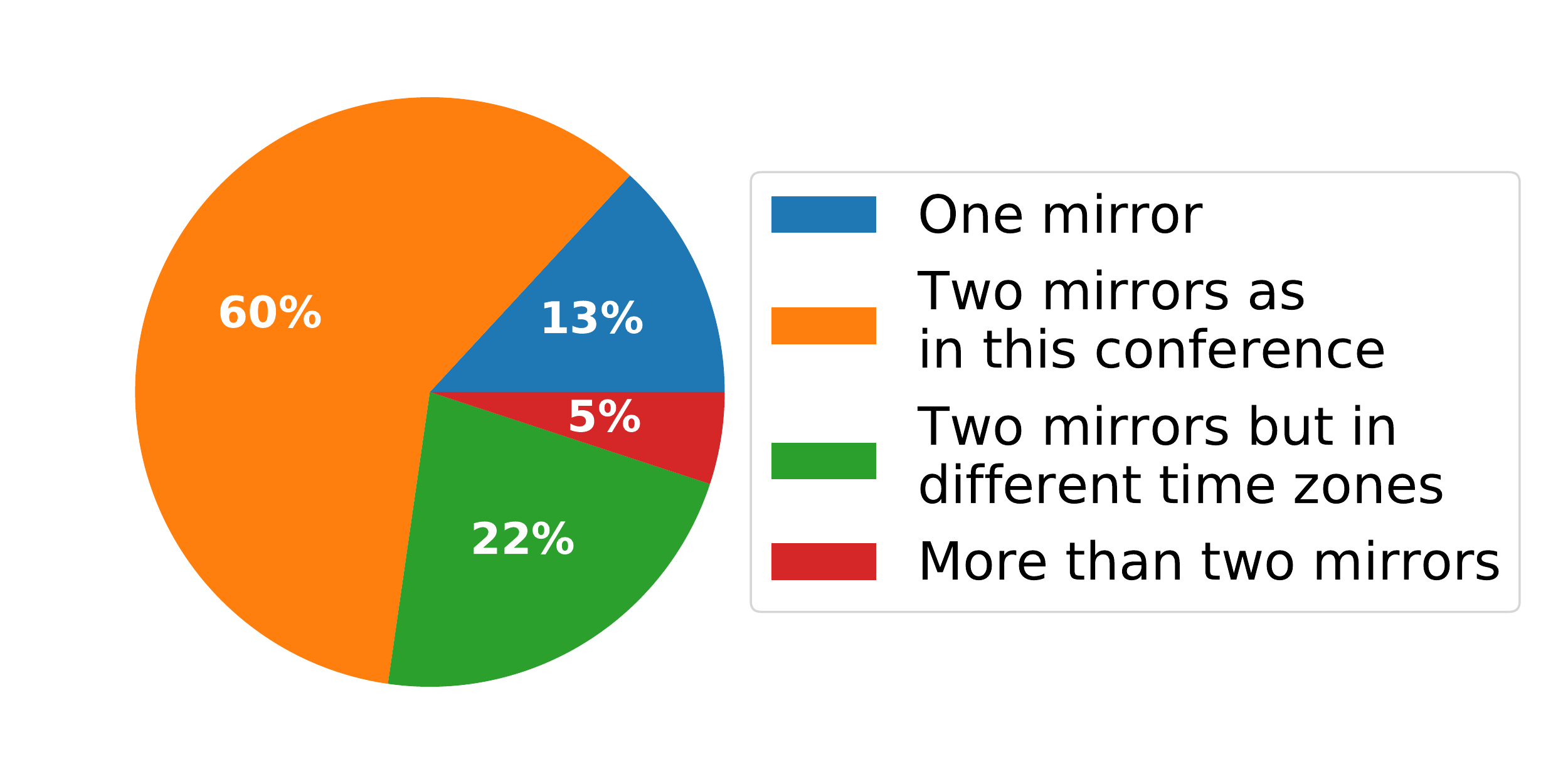}
	\caption{Overview of expected number of introduced mirrors.}
	\label{figure:overview_introduced_mirror}
\end{figure}

\begin{figure}[!h]
	\centering
	\includegraphics[width=12cm]{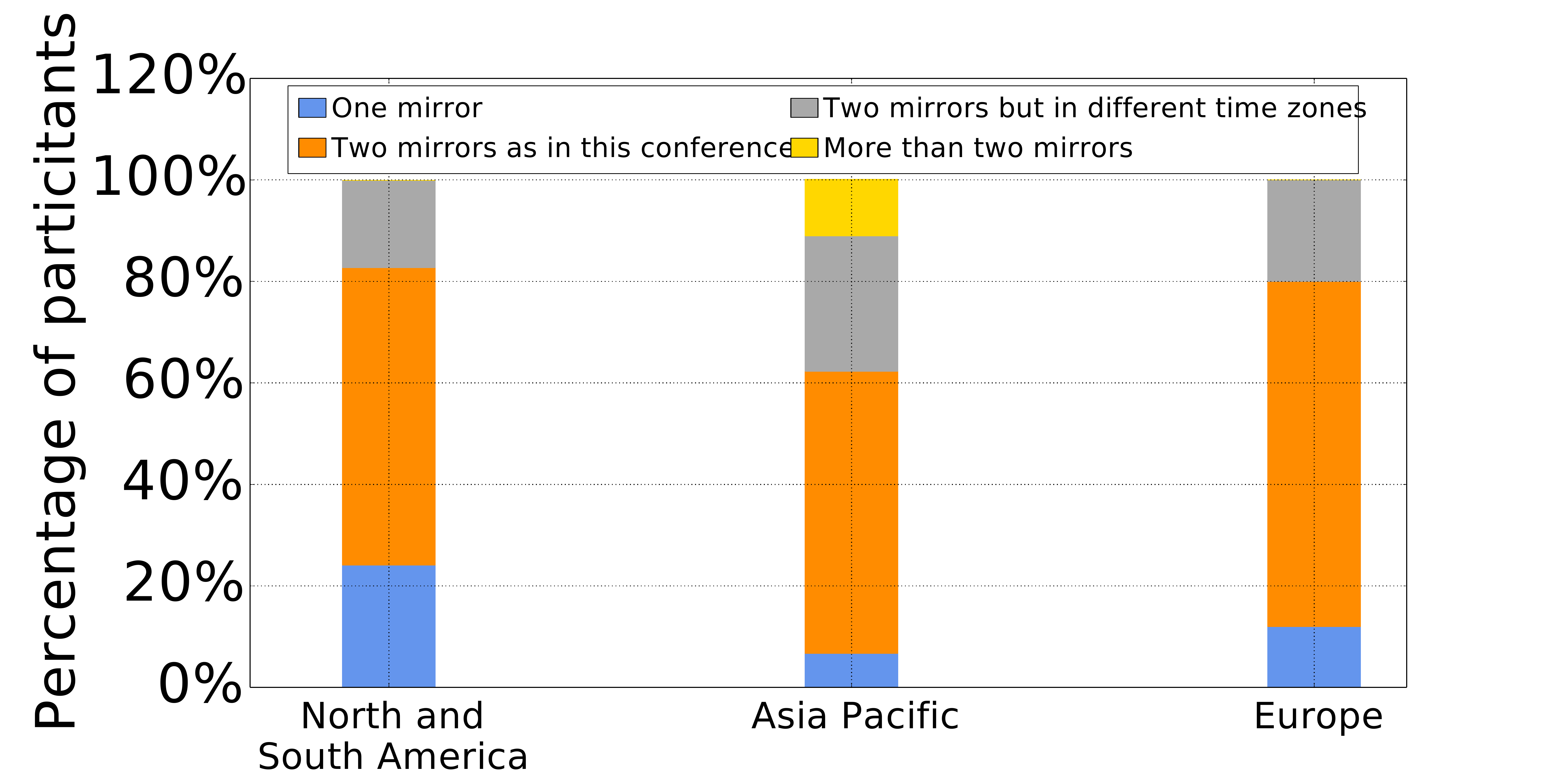}
	\caption{Expected introduced mirror of participants in different regions.}
	\label{figure:introduce_mirror_region}
\end{figure}

\begin{figure}[!h]
	\centering
	\includegraphics[width=12cm]{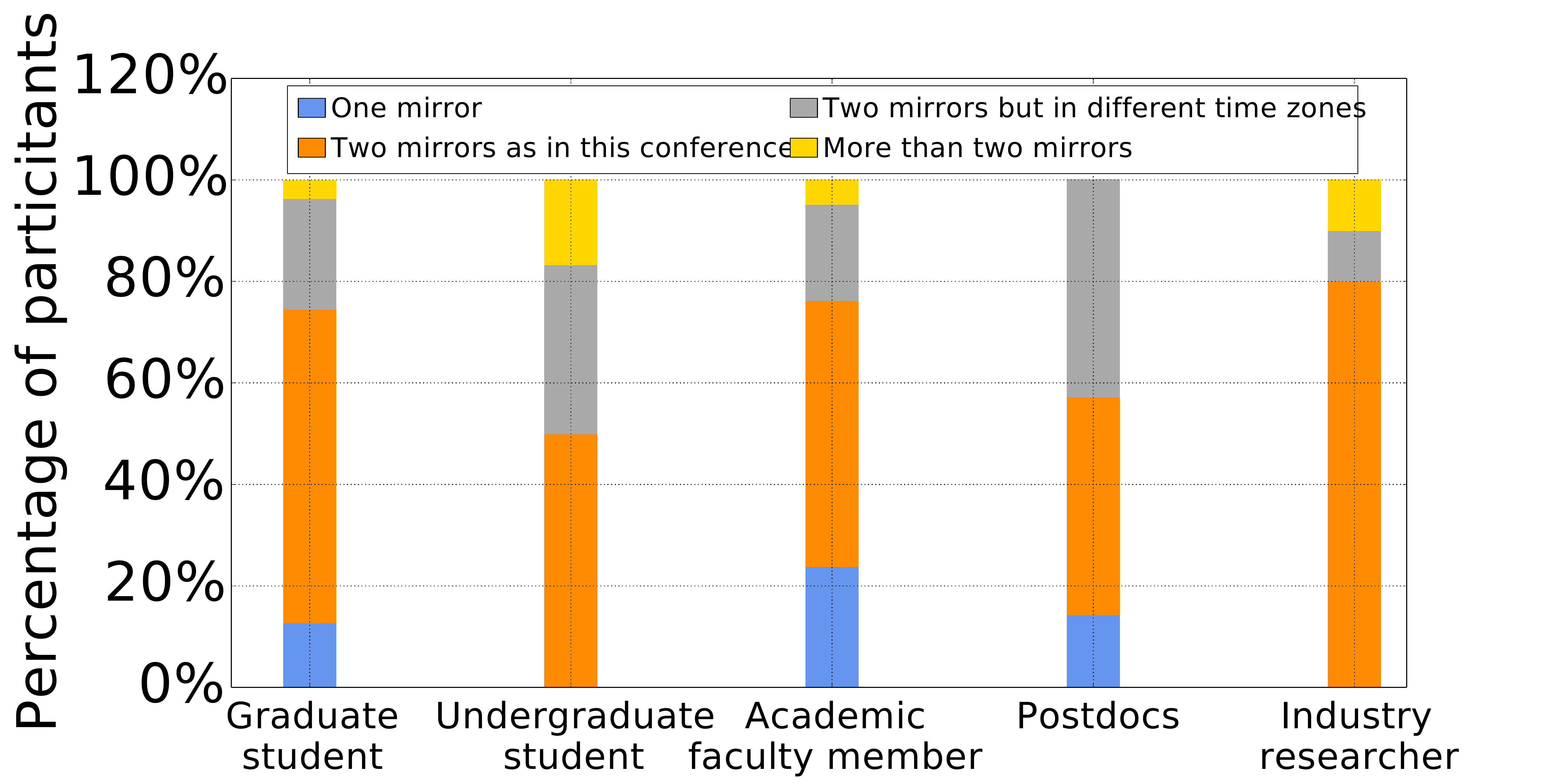}
	\caption{Expected introduced mirror of participants in different professions.}
	\label{figure:introduce_mirror_profession}
\end{figure}




In our survey, we asked how many mirrors are preferred. Fig. \ref{figure:overview_introduced_mirror} shows the results: 60\% prefer two mirrors in the current time zone, 22\% prefer two mirrors in different time zones, and 5\% prefer more than two mirrors. We take an in-depth look at the participant distributions in answering this survey question. Fig. \ref{figure:introduce_mirror_region} and Fig. \ref{figure:introduce_mirror_profession} show the distribution according to the region and profession of the participants. They conform to general expectation, where Asia-Pacific participants prefer more mirrors or mirrors in revised time.

\begin{figure}[!h]
	\centering
	\includegraphics[width=12cm]{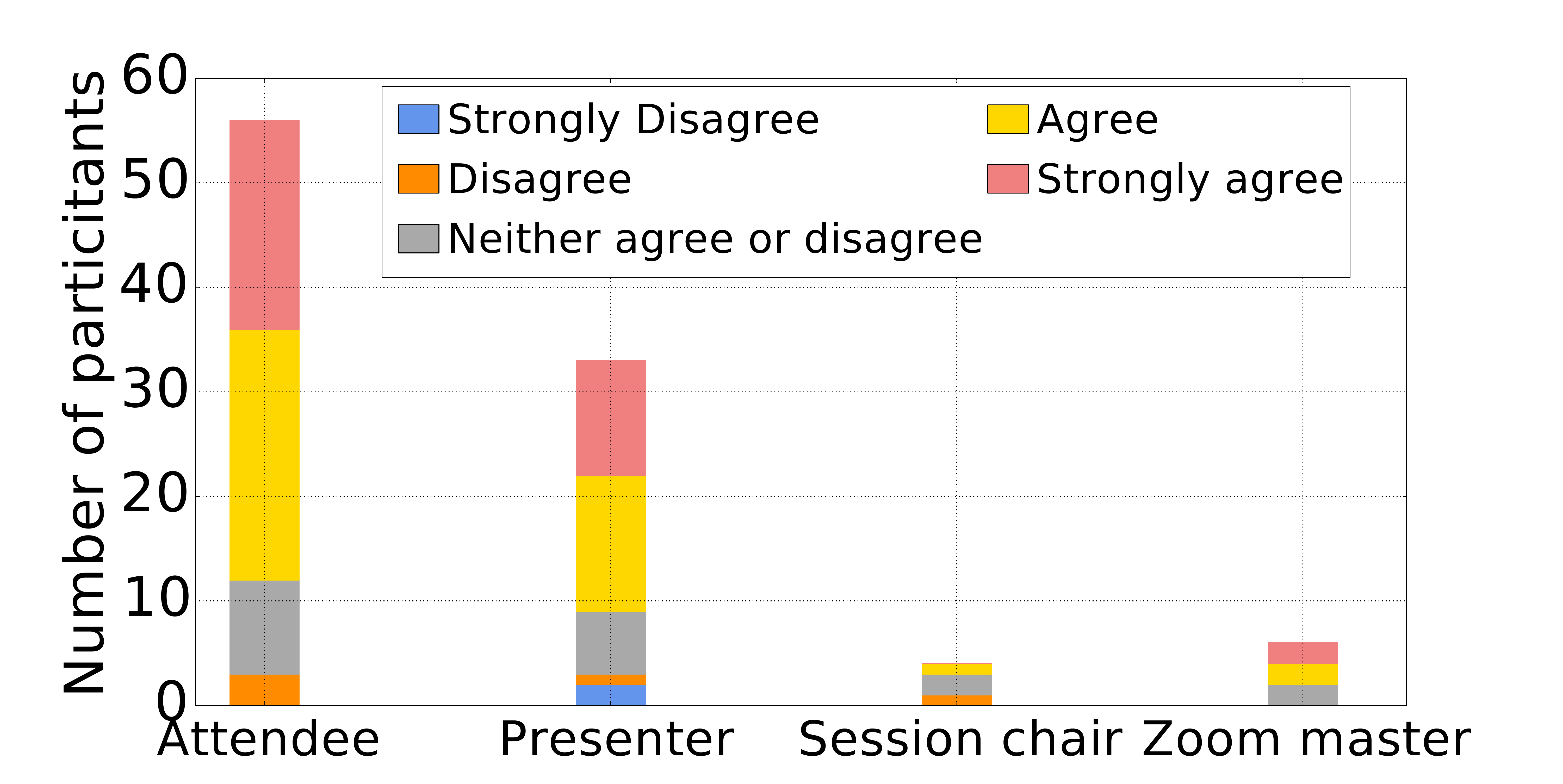}
	\caption{Agreement degree on whether two mirrors are helpful of different roles.}
	\label{figure:mirror_helpful_agreement_role}
\end{figure}

In our survey, we asked whether the mirror program is helpful. Fig. \ref{figure:mirror_helpful_agreement_role} shows strong agreement and agreement by attendees and presenters. The agreement of session chairs and zoom masters is less probably because they partially fall in the organizers role.

\begin{figure}[!h]
	\centering
	\includegraphics[width=12cm]{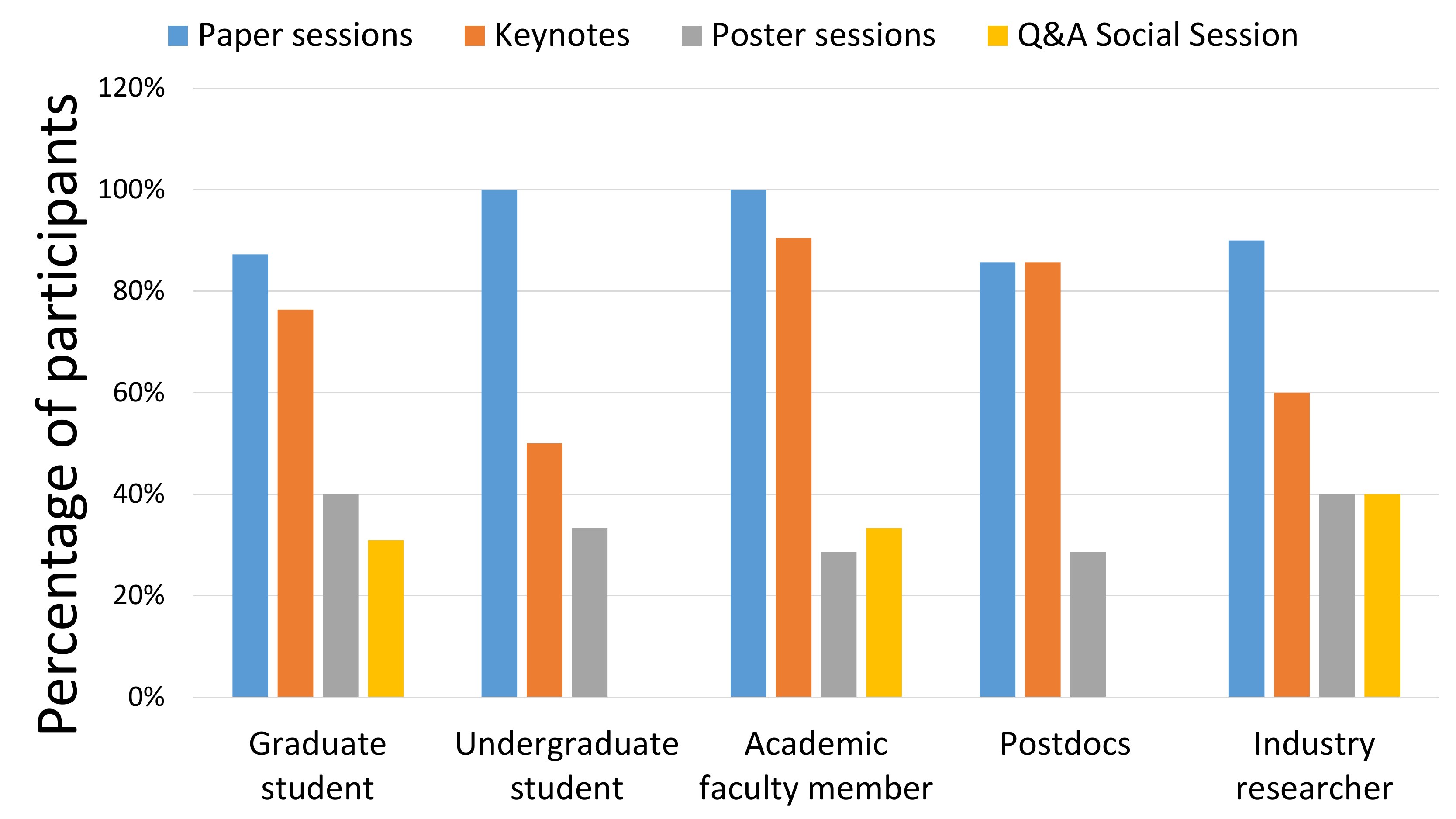}
	\caption{Selection of participants in different profession on session category.}
	\label{figure:session_category_profession}
\end{figure}

\begin{figure}[!h]
	\centering
	\includegraphics[width=12cm]{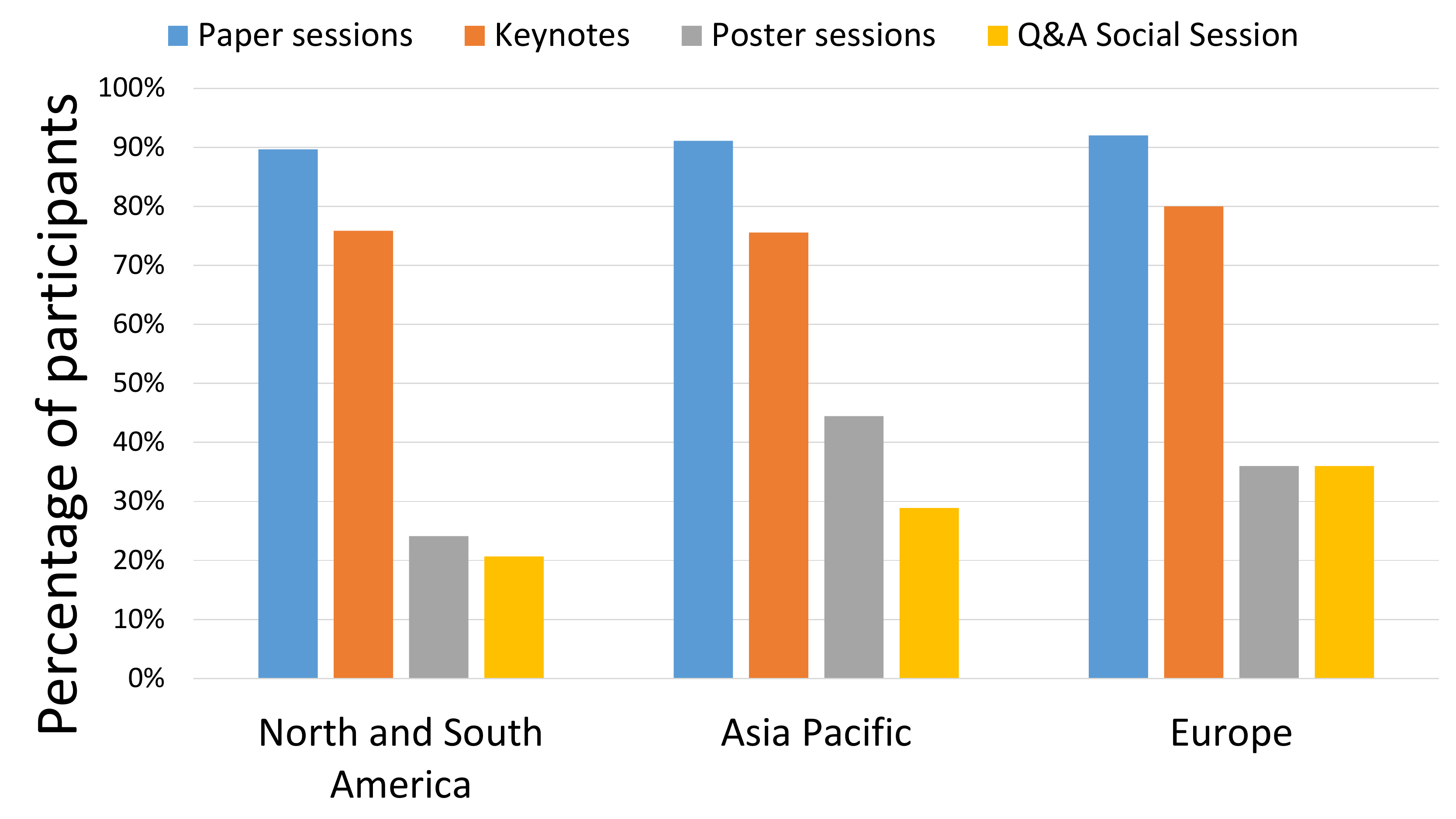}
	\caption{Selection of participants in different region on session category.}
	\label{figure:session_category_region}
\end{figure}

\begin{figure}[!h]
	\centering
	\includegraphics[width=12cm]{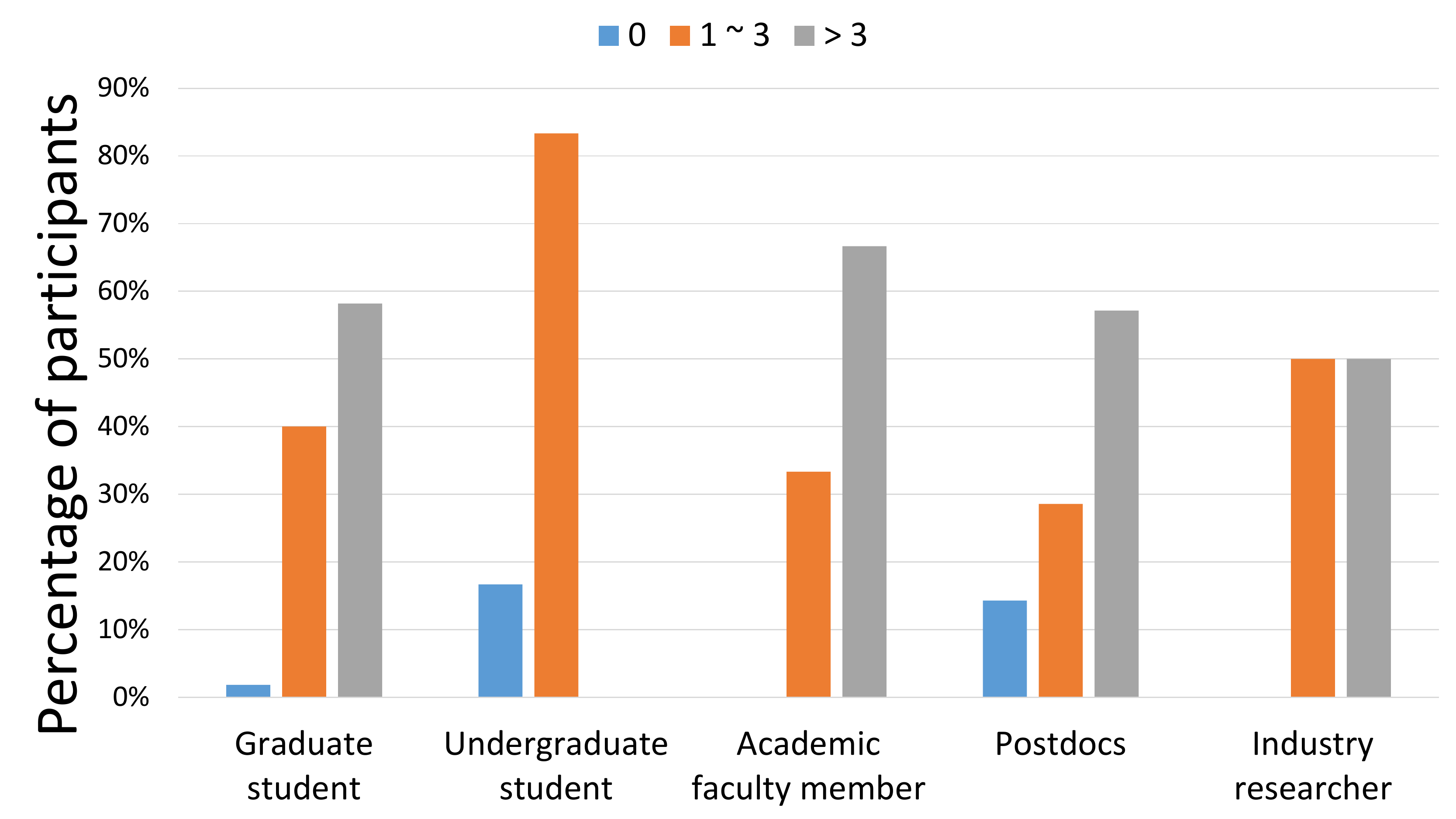}
	\caption{Selection of participants in different profession on technical paper session.}
	\label{figure:technical_session_profession}
\end{figure}

\begin{figure}[!h]
	\centering
	\includegraphics[width=12cm]{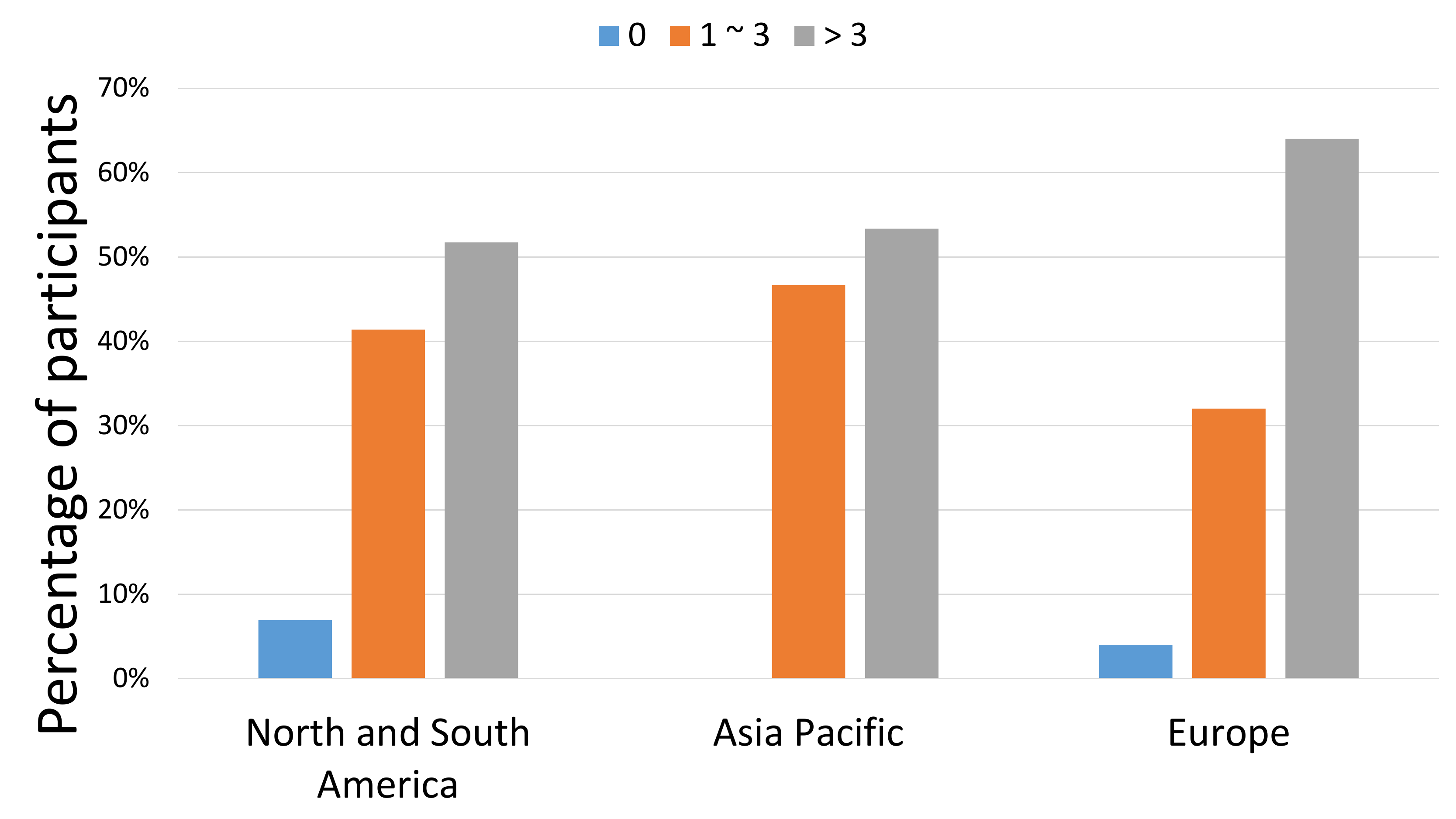}
	\caption{Selection of participants in different region on technical paper session.}
	\label{figure:technical_session_region}
\end{figure}

ACM e-Energy 2020 has four session categories: the technical paper sessions, keynotes, poster sessions and social sessions. We asked which session category people participated in (multiple choices). Fig. \ref{figure:session_category_profession} and Fig. \ref{figure:session_category_region} show the results according to attendees' professions and regions respectively. We see that technical paper sessions are more attractive than keynotes. For example, 100\% of academic faculty members have gone to technical paper sessions, and 92\% have gone to keynotes; which means that 8\% have not attended any of our three keynotes. The difference is more significant for undergraduate students and industry researchers; 100\% undergraduate students have gone to the paper session, yet only half of them have gone to the keynotes. This is probably because that ACM e-Energy 2020 has diverse topic of interest and undergraduate students if the keynotes fall out of the interest of undergraduate students, they may not attend. We can also see that no undergraduate student attended the social session.

ACM e-Energy 2020 has nine technical paper sessions. In our survey, we asked how many technical paper sessions the attendees have gone to. Fig. \ref{figure:technical_session_profession} and Fig. \ref{figure:technical_session_region} show the results according to attendees' professions and regions respectively. We see that academic faculty members went to more sessions, e.g., 68\% have gone to more than three technical paper sessions. Graduate students and postdocs come next.
As a comparison, no undergraduate student has gone to three technical sessions or more. From the viewpoint of regions, attendees based in Europe went to more technical sessions than those based in Asia Pacific or America.

\section{Discussions and Conclusion}
\label{conclusion}

At the time we decided to hold ACM e-Energy virtually, we did not have any prior experience organizing a virtual conference. Below, we summarize a few key takeaway points and highlight aspects that have some room for improvement.

\begin{itemize}

\item In the ACM e-Energy program, we reserved lunch sessions. This turned out to be unnecessary since it is impossible to synchronize lunch time across that many time zones. It is though important to have breaks to reduce the monotonous activity of attending sessions.

\item Two mirrors may partially overlap. There are sessions that are more popular, e.g., keynotes, a session with best papers, etc. It is necessary to ensure that these sessions will not overlap with others. In ACM e-Energy 2020, the last session in the London mirror partially overlapped (two out of four papers in this session) with the keynotes in the PDT mirror. We observed that attendees switched mirrors to watch the keynote. In MACS, we can add a constraint on some sessions by setting them as non-overlapping.

\item Zoom cannot produce videos when the Zoom session is on-going and it takes a long time to generate videos. Therefore, the conference should not plan to record a session in an early mirror and broadcast it in a later mirror. One alternative is to use a zoom meeting for each session. This is at the expense of making it more complex to attend sessions as new log on links are necessary. 

\item The most popular social session was the poster session when we sent attendees to the break out rooms. We learned two important lessons: 1) a social session should have some topics to start the conversation. A poster can serve the topic to start conversation; and 2) attendees should be able to move among break out rooms by themselves. To move among break out rooms, the attendees need to have co-host status in bulk. Unfortunately, Zoom does not provide a method to assign all attendees co-host status. In our poster session, we had multiple Zoom masters to change the attendees' status one-by-one yet this obviously does not scale.

\item We believe that virtual conferences are adept at communicating the technical content well via interactive talks and Q\&A. However, social interaction using current technology is somewhat limited since face-to-face discussion has an edge compared to virtual meetups. It would be great for tools such as Zoom and Slack to provide enhanced technology that can significantly reduce the social interaction barrier. Till such time, perhaps conferences can alternate between physical and virtual; after all reducing the carbon footprint by 50\% is still an enormous benefit.

\end{itemize}

At the time that this paper is written, there is no sign that COVID-19 pandemic is finished, and dearly indications are such that pandemics more generally cannot be eliminated. This paper contributes to the literature as we present experience of a conference with mirrored program design. We believe our experiences to be helpful in planning future virtual, or hybrid conferences.


\bibliographystyle{plain}
\bibliography{ref}

\end{document}